\documentclass[doublecol]{epl2} 
\usepackage{amsmath,amssymb}
\usepackage{bm,stmaryrd}

\title{Odd-even effect in the dominant order of self-assembly of cross junctions in space dimension \mth{d\ge 3}}
\shorttitle{Self-Assembly of Hyper-Cross Junctions in General Dimensions} 

\author{Kazuya Saito\inst{1,2}}
\shortauthor{K. Saito}

\institute{                    
 \inst{1} 
Center for Computational Sciences, University of Tsukuba, Tsukuba, Ibaraki 305-8571, Japan

\inst{2} 
Research Center for Thermal and Entropic Science,
Graduate School of Science, The University of Osaka, Toyonaka, Osaka 560-0043, Japan
}

\abstract{
We consider the self-assembly of cross junctions in a general space dimension ($d$) 
as an extension of the problem studied in a previous paper for $d = 3$.  
This problem is equivalent to constructing a $d$-dimensional hypercubic jungle gym,
at all junctions of which $2d$ rods with different colours meet.
The analysis reveals a unique feature of the $d = 3$ case: 
the forced presence of at least one perfectly-ordered (singly coloured) direction (axis),
in contrast to the possible absence of such a direction in $d\ge 4$. 
However, we will show that the uniaxial order is overwhelming not only in $d=3$ but also 
for $4\le d \le 7$ and odd $d\ge 9$ 
in a sufficiently large system. 
For even $d\ge 8$, isotropic states dominate, leading to the alternation of dominant states
between the uniaxial and isotropic orders depending on the parity of $d \ge 7$.}

\begin{document}

\maketitle

\section{Introduction}
The selection of a limited domain from all possible configurations in a thermodynamic ensemble 
consisting of a vast number of micro entities is 
an example of symmetry breaking \cite{critical1,yn,critical3}.
This has offered fascinating issues \cite{baxter}.
While the selection by the energetic effect is often manageable and aligning with intuitions,
that driven by entropy is complicated.
The structure of aggregates of rigid spheres formed by the so-called Alder transition can be 
regarded as an example \cite{HS1,HS3,HS4,HS6,HS7}.
The order-by-disorder concept \cite{ObyD} and phenomena \cite{exMS,exMS_4d,ObD_prb} are also remarkable.
This Letter addresses such an issue. 
The possibility of reaching a universal conclusion about the emerging order 
in thermodynamic ensembles is discussed in a general space dimension of $d\ge 3$.

\begin{figure}[tb]
\begin{center}
\includegraphics[width=7cm]{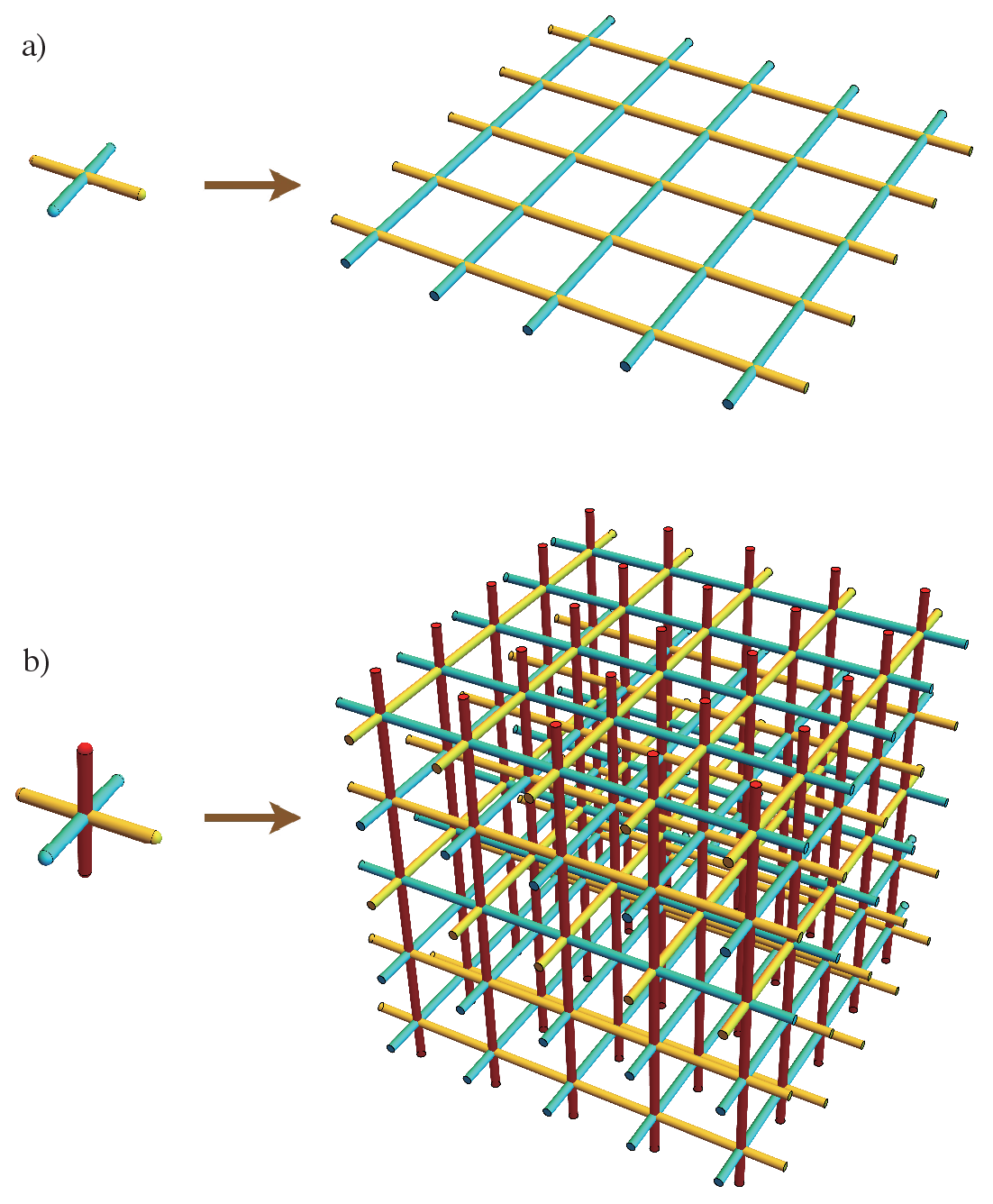}
\end{center}
\caption{Cross junctions and example of their self-assembled states in dimension two (a) and three (b). While there is no variety in dimension two except for exchanging colours,  many possibilities exist 
in dimension three.}
\label{junctions}
\end{figure}

The micro entity in question is named a cross junction,
while referring to cross-polytopes \cite{polyhedron}.
That of dimension $d$ consists of $d$ segments with a unit length.
Each segment possesses a different colour from the rest. 
The segments intersect at right angles at their centres.
Figure \ref{junctions} illustrates examples of cross junctions in dimensions two and three.
Upon their self-assembly, 
we assume that 
a segment from one junction joins linearly with a segment of the same colour from another junction. 
Figure \ref{junctions} also illustrates examples of self-assembled ensembles.
For $d\ge 3$, the self-assembled ensemble is a (hyper)cubic jungle gym,
where all junctions are cross junctions, and all straight lines are painted a single colour.

While the self-assembled states are trivial for dimension two,
they are not in dimension three.
In the example for $d=3$ in fig.\ \ref{junctions}, all lines along the vertical axis are coloured red,
whilst the orientations of square nets (with two colours) parallel to the basal plane are irregular.

We crudely characterise self-assembly by the number of directions (axes) 
along which all lines share the same colour. 
The state where all parallel lines have the same colour is termed ``completely ordered'' in this Letter.
Hereafter, we refer to states with $n$ completely ordered directions as the $\langle n\rangle$-state.
If helpful, we indicate its dimension $d$ with a suffix as a $\langle n\rangle_d$-state.

A previous paper \cite{jcp} proved that 
the presence of at least one completely ordered direction (axis) is unavoidable in $d=3$.
Also discussed was that such an uniaxial order, i.e., $\langle 1\rangle$-states,  dominates in thermodynamic ensembles.
The purpose of this study is to place such characteristics of $d=3$ in a broader context
of similar problems in general dimensions $d\in \mathbb{N}$ (positive integer).
Interestingly, we can plausibly make a statement: 
the uniaxial order is most probable for general odd $d\ge 3$.
For even $d$, the isotropic states dominate except for $d=2,4,6$,
where uniaxial states prevail.
This conclusion is probabilistic, but with a confidence resembling that of thermodynamics. 
Therefore, it can be termed entropic.


Despite having no particular applications to problems in the real world of dimension three, 
an alternative formulation of the discussed issue can be inspiring. 
First, consider a ``quantised'' space of dimension $d$, 
where the quanta are a hypercube of dimension $d$ (with $2d$ faces). 
Each pair of parallel faces has a distinct colour from the others. 
The conclusion we will reach states that a single dimension of this space of dimension of odd $d$ is special despite the isotropic nature of hypercubic quanta,
if the space is composed of a vast number of quanta combined via faces of the same colour. 
The second formulation is mapping onto the Potts model \cite{Potts,PottsModel}. 
When we assign a spin variable with $d$ choices (colours) to each face 
and assume an antiferroic interaction that hates the same colours between edge-sharing faces, 
our quantised space (or jungle gym) of dimension $d$ is naturally mapped to the ground state of the antiferroic $d$-states Potts model. 
Therefore, the present problem is to identify a universal property of the ground states of the Potts model on a certain class of lattice geometry, applicable across space dimensions.

\section{Setting}
We denote the number of possible $\langle n\rangle_d$-states 
by $v(n;d)$,
which counts states while distinguishing both colouring and orientation with respect to the axes.
Obviously, $v(d;d)=d!$ and $v(d-1;d)=0$.
The latter property arises from colour exclusion at each cross junction.
The sum of all possibilities is expressed as
\begin{equation}
V_d=\sum_{i=0}^d v(i;d)\label{Wd}
\end{equation}
which strongly depends on the system size $N=L^d$ ($N$ being the total number of junctions).
The behaviour of $v(n;d)$ and $V_d$ at very large $N$, consequently very large $L$, 
is of our interest.
Throughout this Letter, 
we will analyse the problem by first assuming a fixed $L$ and increasing it,
rather than assuming a fixed $N$ and varying $d$.

While enumerating the number of states,
we encounter many terms of the form
\begin{equation}
\prod_{i=0}^{d-2} \left(a_i\right)^{L^i}
\end{equation}
with $a_i\in \mathbb{N}$.
For the sake of conciseness, we denote the term as 
\begin{equation}
\llbracket a_0,a_1, ..., a_{d-2}\rrbracket
\end{equation}
with the rule that unity ($1$) is assigned when a component is absent.
Adding an arbitrary number of ones at the end, enclosed in double brackets, does not alter the term.
Conversely, the tail of ones can be removed.
When the components of $A$ for $i\ge j$ coincide with those of $Q$,
we may write
\begin{equation}
A=\llbracket a_0, ..., a_{j-1}, \llbracket Q\rrbracket\rrbracket .
\end{equation} 
This notation leads to
\begin{equation}
(Q)^{L^k} = \llbracket \underbrace{1, ..., 1}_{k}, \llbracket Q\rrbracket\rrbracket .
\end{equation}

The notation facilitates judging the relative dominance of two terms.
For $\llbracket B\rrbracket=\llbracket ..., b_n, \llbracket Q\rrbracket\rrbracket$ and 
$\llbracket C\rrbracket=\llbracket ..., c_n, \llbracket Q\rrbracket\rrbracket$,
we have
\begin{equation}
B/C=\frac{\llbracket B\rrbracket}{\llbracket C\rrbracket} =\left\llbracket ..., \frac{b_n}{c_n}\right \rrbracket
\end{equation}
because the two $\llbracket Q\rrbracket$s in the double brackets of $\llbracket B \rrbracket$ and $\llbracket C \rrbracket$ cancel each other out.
Therefore, the value of the last $b_i/c_i$ different from unity within double brackets indicates
which of $B$ or $C$ dominates in the limit of $L\rightarrow \infty$.


\section{Possible absence of a completely ordered axis in $d\ge4$}
The previous paper \cite{jcp} proved 
\begin{eqnarray}
v(0;3)&=&0\label{v03}\\
V_3&=& 9\cdot2^L\\\label{v3}
&=&\llbracket 9,2\rrbracket,\nonumber
\end{eqnarray}
where the factor 9 comes from the choices of the colour and orientation of the unique, i.e.,
the completely ordered axis.
The former means the presence of at least a single completely ordered axis
in $d=3$, while the latter implies the overwhelming possibility of $\langle 1\rangle$-states 
with respect to the $\langle 3\rangle$-states in a large system because of $v(3;3)=6$ 
by the colour permutation.

\begin{figure}[tb]
\begin{center}
\includegraphics[width=6cm]{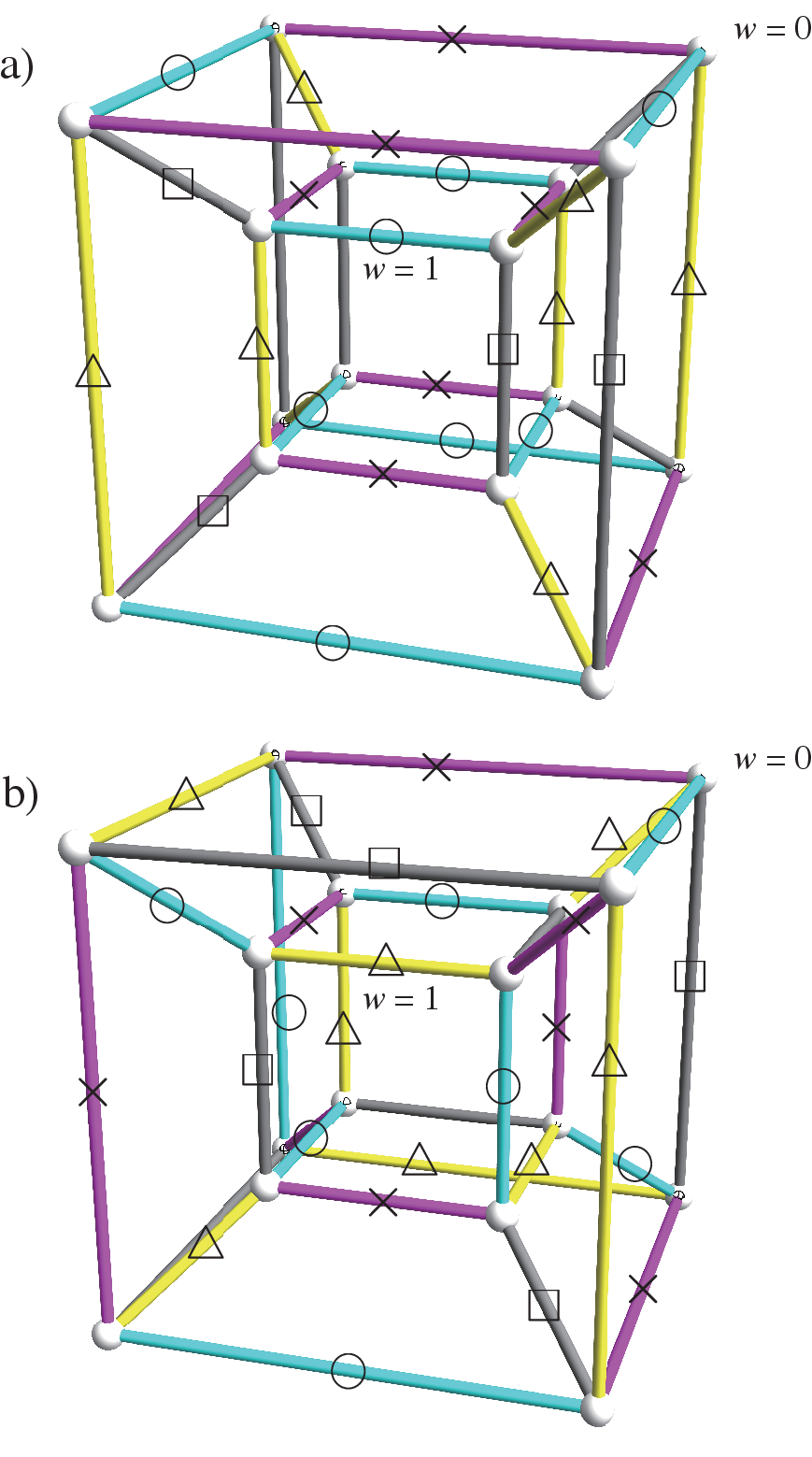}
\end{center}
\caption{Two examples of a $2\times 2\times 2\times 2$ part of self-assembled states without completely ordered axes in dimension four.
a) specified by eq.\ \ref{splitcase},
which decomposes the dimension into $2+2$;
b) all axes are equivalent.
Large (outer) and small (inner) cubes (of dimension three) represent sectors with different coordinates in the fourth dimension ($w$), which is depicted along the body diagonals of the cubes, with the inward direction being positive.
The half number of the arms of each junction, such as those at the outer/inner sides of hypercubes, is omitted for clarity.
Different colours are distinguished using different symbols marked for most segments.}
\label{4d_zero}
\end{figure}

First, we prove $v(0;d)\ne 0$ for $d= 4$.
To this end, an example suffices.
Figure \ref{4d_zero} shows two examples of parts with of $2^4$ junctions.
We see that, at all depicted spheres, which are the centres of cross junctions, 
four arms of different colours meet,
confirming the successful formation of illustrated parts.
Note that a successful construction of parts of this size guarantees
an extension covering the whole space in dimension 4 by putting identical copies
with a shift by two in respective directions (axes). 
Therefore, we have at least two examples of completely self-assembled ensembles
expandable to $L\rightarrow \infty$.

We look at more details.
The colours of segments are designated as $\{c_1,c_2,c_3,c_4\}$ according to the direction of 
the axes ($i = 1 - 4$) of a cross junction.
The example in fig.\ \ref{4d_zero}a is constructed as follows:
Using the abbreviations E and O for even and odd, respectively,
we set the colours by 
\begin{multline}
\{c_1,c_2,c_3,c_4\}\\
=
\begin{cases}
\{\text{C},\text{M},\text{Y},\text{K}\} & \tx{if\ }(n_1+n_2,n_3+n_4)\tx{\ is\ }(\tx{E},\tx{E})\\
\{\text{C},\text{M},\text{K},\text{Y}\} & \tx{if\ }(n_1+n_2,n_3+n_4)\tx{\ is\ }(\tx{O},\tx{E})\\
\{\text{M},\text{C},\text{Y},\text{K}\} & \tx{if\ }(n_1+n_2,n_3+n_4)\tx{\ is\ }(\tx{E},\tx{O})\\
\{\text{M},\text{C},\text{K},\text{Y}\} & \tx{if\ }(n_1+n_2,n_3+n_4)\tx{\ is\ }(\tx{O},\tx{O})
\end{cases}\label{splitcase}
\end{multline}
%
for the junction at $(n_1,n_2,n_3,n_4)$.
Here, we used C, M, Y, and K as dummy colour symbols.
All meshes parallel to the basal (e.g., $xy$) plane contain only two colours (C and M) and stack alternately in their orientation along both of the other axes ($z$ and $w$).
A similar situation holds for the meshes parallel to the $zw$-plane with a different combination of two colours (Y and K).
In contrast, squares on all of the other planes contain four colours. 

Although the stacking of two-colour meshes is regular (alternate) in both split pairs in fig.\ \ref{4d_zero}a,
the way of stacking does not affect the incompleteness of the order.
For example, the number of colours gathering at the corner spheres in the uppermost square
remains unchanged at four if the square is rotated by $\pi/2$.
The same applies to any squares with edges of two colours.
Therefore, the split case can be regarded as an example
of similar patterns.

In contrast to the split case of fig.\ \ref{4d_zero}a, 
all squares contain four colours in fig.\ \ref{4d_zero}b.
More precisely,
all groups of colour segments are mutually equivalent to each other.
In other words, they are ``symmetric.''
In this case, the variety in colour rather reduces the possible ways of self-assembly.
It is difficult to imagine deriving a variety of other units from fig.\ \ref{4d_zero}b.
At present, the enumeration of the states is incomplete.
Indeed, we can imagine ``asymmetric'' hypercubes.
Nevertheless, we plausibly assume that such ensembles will not surpass those constructed systematically,
because the rule of self-assembly restricts the possibilities, causing a strong correlation
in junctions' orientations.

We proceed with the possibility of $v(0;d)$ for $d\ge 5$.
The way to get a variety of ensembles without a completely ordered axis in $d=4$
relies on the orientational freedom of both split sectors (of dimensions $(d-2)$ and $2$).
Indeed, its absence for a string, corresponding to the $d=1$ case, explains eq.\ \ref{v03}.
Starting from the ensemble in dimension $(d-2)$,
we can similarly construct $\langle 0\rangle_d$-states,
as will be described in more detail later.
Therefore, the following holds:
\begin{equation}
v(0;d)\ne 0
\end{equation}
for a general dimension $d\ge 4$. 
Therefore, eq.\ \ref{v03} is a unique property of $d=3$.

\section{Dominant states in $d\ge 4$}
We consider the issue of what is the most probable number of completely ordered axes in
general dimensions. 
In other words, what is the distribution of $\langle n\rangle$-states depending on the space dimension $d$?

We consider the increase in the dimension by one: $(d-1) \rightarrow d$.
When $L$ copies of a $\langle n-1\rangle_{d-1}$-state are stacked along the newly added dimension,
the resultant ensemble can be regarded as a $\langle n\rangle_d$-state.
Therefore, at least, 
there are the same number of $\langle n\rangle_d$-states.

On the other hand,
we always obtain a $\langle 1\rangle_d$-state
if we pick up $L$ self-assembled ensembles of dimension $(d-1)$ at random
and stack them along the new axis.
Although the resultant ensembles in dimension $d$ contain those with $n>1$, such as $n=d$,
they are always minor in comparison with those of $n=1$, as discussed below.
Therefore, the number of $\langle 1\rangle_d$-states ($d\ge 3$) is 
\begin{eqnarray}
v(1;d)&\sim& d^2\cdot V_{d-1}^L\label{wd_1}\\
&=&\llbracket d^2,\llbracket V_{d-1}\rrbracket\rrbracket\nonumber
\end{eqnarray}
including the possibility arising from $\langle 0\rangle_{d-1}$-states.
$v(1;3)$ fulfils this formula.
Here, $d^2$ in eq.\ \ref{wd_1} comes from the choice of a colour and an orientation of the unique axis. 
Since any $\langle 1\rangle_d$-state can be disassembled into $L$ pieces of 
$(d-1)$-dimensional assemblies and parallel strings embedded in the particular dimension,
the present construction of $\langle 1\rangle$-states encompasses all possible ones.

Similarly, we obtain $\langle 2\rangle_d$-states as many as
\begin{equation}
d^2\cdot(d-1)^2\cdot\left[\frac{V_{d-1}-v(0;d-1)}{(d-1)^2}\right]^L,\label{wd_2}
\end{equation}
where 
the denominator inside the bracket indicates the need for at least a perfect axis, 
whereas the numerator chooses a colour and an orientation, 
in addition to a pre-factor arising from the choices of colours and orientations.
Obviously, eq.\ \ref{wd_2} is smaller than and negligible compared with eq.\ \ref{wd_1}
in the large $L$ limit.
The same applies to all $\langle n\rangle_d$-states up to $n=d-1$,
which is $v(d-1;d)=0$.
Therefore, we reach 
\begin{equation}
V_d\sim v(1;d)+v(0;d).\label{Vd_01}
\end{equation}
Here, we retain $v(0;d)$ because we have not estimated it.

We begin enumerating $v(0;d)$.
Adding a new dimension considered above is inapplicable to this purpose.
Only the systematic construction we know is the case of fig.\ \ref{4d_zero}a.
We adopt a similar construction.
To obtain a self-assembly consisting of $N$ ($=L^d$) cross junctions using
self-assemblies of dimension $(d-2)$,
we need $L^2$ pieces of such an assembly.
Thus, the choice of an assembly in dimension $(d-2)$ and its orientation and colouring
yields the total number of  
$\binom{d}{2}^2\cdot (V_{d-2})^{L^2}$. 
Successive additions of new dimensions twice yield a $\langle 2\rangle_d$-state
through the process of $L^2\langle k\rangle_{d-2}$-states ($0\le k\le d-2$) 
$\rightarrow L\langle 1\rangle_{d-1}$-states\ $\rightarrow$  a single $\langle 2\rangle_d$-state.
In this final state, the two newly added dimensions result in 
stacks of $L$ meshes in each of $(d-2)$ directions.
Since their dimensions are independent of the $(d-2)$ dimensions that previously existed,
each mesh can change its orientation, as in fig.\ \ref{4d_zero}a. 
The number of their possible stackings is $2^{(d-2)L}$.
On this background, we define the following quantity for $m\le d/2$: 
\begin{eqnarray}
u(m;d)&=& \binom{d}{m}^2\cdot (V_{d-m})^{L^m}\cdot (m!)^{(d-m)L},\label{ud}\\
&=&\left\llbracket \binom{d}{m}^2, (m!)^{d-m}, \underbrace{1, ..., 1}_{m-2},\llbracket V_{d-m}\rrbracket\right\rrbracket\nonumber
\end{eqnarray}
which is the number of $\langle 0\rangle_d$-states 
constructed from the self-assembled ensembles in dimension $(d-m)$. 
Although we do not have a reliable estimate of additional cases to $u(0;d)$ 
arising from other constructions like fig.\ \ref{4d_zero}b,
it contributes only additively and will be small.
Thus, we suspect
\begin{equation}
v(0;d)\sim \sum_{m=2}^{\lfloor d/2\rfloor} u(m;d).\label{umd}
\end{equation}
Although the case of $d=4$ is straightforward, as it involves only $u(2;4)$,
we treat it together with other cases to avoid repetition.

We now show, for $4\le d\le 7$,
\begin{equation}
V_d \sim v(1;d),\label{main}
\end{equation}
which is valid for $V_3$.
If eq.\ \ref{main} holds, we have
the recursion relation
\begin{eqnarray}
V_d&\sim&d^2\cdot V_{d-1}^L.\nonumber\\
&=&\llbracket d^2,\llbracket V_{d-1}\rrbracket\rrbracket.\label{recursion}
\end{eqnarray}
Then, we see
\begin{eqnarray}
u(2,d)&=&\left\llbracket \binom{d}{2},2^{d-2},\llbracket V_{d-2}\rrbracket\right\rrbracket\label{u2d2}\\
&=&\left\llbracket \binom{d}{2},2^{d-2},(d-2)^2,\llbracket V_{d-3}\rrbracket\right\rrbracket\nonumber\\
u(3;d)&=&\left\llbracket \binom{d}{3},2^{d-3},1,\llbracket V_{d-3}\rrbracket\right\rrbracket.\label{u3d}
\end{eqnarray}
This comparison indicates that they share components of the power higher than $L^3$, but 
$u(2,d)$ has the component of the power $L^2$, in contrast to the absence in $u(3,d)$.
This is also the case for larger $m\le d/2$ available.
Therefore, we reach
\begin{equation}
v(0;d)\sim u(2;d)\label{u2d}
\end{equation}
for $d\ge 5$.

Comparing $v(1;d)$ and $u(2;d)$ indicates which of $\langle 0\rangle$-state 
or $\langle 1\rangle$-state dominates.
Because of eq.\ \ref{u2d2} and
\begin{equation}
v(1;d)=\llbracket d^2,(d-1)^2,\llbracket V_{d-2}\rrbracket\rrbracket,
\end{equation}
the second component is essential.
That of the $\langle 1\rangle$-state, $(d-1)^2$, is greater than $2^{d-2}$, that of the $\langle 0\rangle$-state for $d\le 7$.
Namely, eq.\ \ref{main} stands in this range of $d$.
Therefore, for $4\le d\le 7$, thermodynamic ensembles of cross junctions exhibit
the $\langle 1\rangle$-states despite the possibility of $\langle 0\rangle$-states.

\begin{table*}
\caption{\label{tab:tableW} Components of possible numbers calculated successively in increasing the space dimension ($d$).
The decisive components are shown in bold.
For $d\ge 7$, the set of components, including the bold one, constitutes the asymptotic $V_d$.}
\begin{center}
\begin{tabular}{cccccccccc}
\hline
space dimension & \multicolumn{4}{c}{$\langle 1\rangle$-states}&&\multicolumn{4}{c}{$\langle 0\rangle$-states}\\
 ($d$)&$a_0$&$a_1$&$a_2$ &$a_{\ge 3}$&&$a_0$&$a_1$&$a_2$&$a_{\ge3}$\\
\hline
$6$&$36$&$25$&$16$&$9$\\
$7$&$49$&$\bm{36}$&$25$&$\llbracket V_4\rrbracket $ &&$441$&$32$&$25$&$\llbracket V_4\rrbracket $\\
$8$&$64$&$49$&$\llbracket V_6\rrbracket$& &&$784$&$\bm{64}$&$\llbracket V_6\rrbracket$\\
$9$&$81$&$784$&$\bm{64}$&$\llbracket V_6\rrbracket$&&$1296$&$128$&$49$&$\llbracket V_6\rrbracket$\\
$10$&$100$& $81$&$\llbracket V_8\rrbracket$& &&$2025$&$\bm{256}$&$\llbracket V_8\rrbracket$\\
$11$&$121$ & $2025$ & $\bm{256}$&$\llbracket V_8\rrbracket$&& $3025$&$512$&$81$&$\llbracket V_8\rrbracket$\\
\\
even $d\ge 8$ & $d^2$ &$(d-1)^2$ &$\llbracket V_{d-2}\rrbracket$& &&$\binom{d}{2}^2$& $\bm{2^{d-2}}$ & $\llbracket V_{d-2}\rrbracket$ \\
odd $d\ge 9$ & $d^2$ & $\binom{d-1}{2}^2$& $\bm{2^{d-3}}$ & $\llbracket V_{d-3}\rrbracket$&&$\binom{d}{2}^2$ & $2^{d-2}$ & $(d-2)^2$& $\llbracket V_{d-3}\rrbracket$\\
\hline
\end{tabular}
\end{center}
\end{table*}

Since the above comparison of $v(1;d)$ and $u(2;d)$ is meaningful up to $d-2=7$,
we can say that the $\langle 0\rangle$-state is overwhelming for $d=8$, but 
the dominant state goes back to the $\langle 1\rangle$-state for $d=9$.
Note that the majority of $\langle 1\rangle_9$-states are derived from $V_8$, 
where the majority are $\langle 0\rangle$-states.

Proceeding further is not straightforward 
due to the possible mixed variation of dominance of these states.
Therefore, we successively constructed the main term, $V_d$, based on $V_{d-1}$ ($\langle 1\rangle$-state)
or $V_{d-2}$ ($\langle 0\rangle$-state) as tabulated in Table \ref{tab:tableW}.
Interestingly, the alternation of the dominant state extends for $d\ge 10$.
We identify the ``rule'' in each group for the construction of the more dominant term,
for which the components are indicated in the last lines of Table \ref{tab:tableW}.
The dominance of $u(2;d)$ among $u(m;d)$ was confirmed even for $d\ge18$,
above which all constructions of $\langle 0\rangle$-states (from $\langle n\rangle_{d-m}$-states) require the ensembles in $d\ge 9$, 
the threshold for the state alternation.
The rule in Table \ref{tab:tableW} is certainly consistent with the dominance of $u(2;d)$ among $u(m,d)$ and, consequently, the alternation of the dominant state.
This is because
the exponential dependence is stronger than the quadratic dependence for large $d$.
Therefore, we conclude that the alternation continues without the upper limit in $d$.

\section{Summary and outlook}
Inspired by the recent identification of a counterintuitive conclusion
that octahedral junctions, the cross junction in dimension three, 
self-assemble into a uniaxially ordered state \cite{jcp},
we examined the issue of self-assembly of cross junctions in general dimensions. 

The analysis showed that the forced presence of (at least) one completely ordered axis is
a unique property of the $d=3$ case, arising from the impossibility of dividing $d$ into
a sum of multiple integers larger than $1$.

For $d\ge 4$, $\langle 0\rangle$-state  is possible.
In $d=4$, two distinct groups of such $\langle 0\rangle$-states were identified:
two couples of two colours are separately embedded, or four colours run in every direction.
Based on a naive construction of $\langle 0\rangle$-states starting from $d\ge 3$ by adding dimension two,
which mimics the first group of $d=4$,
the dominant states are identified for $d\ge 4$.
While the uniaxial $\langle1\rangle$-states are overwhelming for $4\le d\le 7$
despite the possible presence of $\langle 0\rangle$-states,
the dominant state alternates for $d\ge 8$.

It is open whether colour-splitting into groups is necessary for $\langle 0\rangle$-states at $d\ge 5$.
A possibility of constructing a variety of ensembles other than the assumed method
cannot be ruled out.
However, such construction contributes to $v(n;d)$ only additively,
being harmless for the present discussion, 
which is interested only in the largest group of self-assembled states.
Note that a symmetrically mixed $\langle 0\rangle$-state,
such as one shown in fig.\ \ref{4d_zero}b for $d=4$ ,
is possible only for a very limited $d$
because the number of edges of a hypercube in dimension $(d-1)$, $2^{d-2}\cdot (d-1)$, 
must be divisible by $d$ for success.

Finally, an emphasis is placed on the fact 
that only completely self-assembled states are within the scope of this Letter.
In general, 
the population of the ordered states is much smaller than that of all possible states, 
as exemplified by an extreme smallness of $2^{-(N-1)}$, 
that of the complete ferromagnetic state of the Ising model. 
This is also the case in the present setting, despite its high degeneracy (though non-thermodynamic). 
In this regard, mapping onto the Potts model described in the Introduction may play an essential role
because it can facilitate reaching a variety of ground states via simulated annealing. 
This approach was applied to the case of dimension three in a previous paper \cite{jcp}, 
which also reported some intriguing properties at finite temperatures in the ordered phase on the course to complete order. 
Preliminary Monte Carlo simulation for dimension four indicated that the transition from the disordered phase to an ordered state occurs at a lower temperature than that in the case of dimension three, contrary to the ferroic case. 
This decreasing trend can be rationalised by a molecular field treatment.
Allowing incompleteness for $d\ge 4$ may introduce an unexpected feature in the approach to complete states, 
as exemplified by \cite{jcp} for the $d=3$ case.







\acknowledgments
This work was supported in part by JSPS KAKENHI in a Grant-in-Aid 
for Transformative Research
Areas ``Materials Science of Meso-Hierarchy'' (JP24H01694).

\end{document}